\documentclass[preprint,showpacs,preprintnumbers,amsmath,amssymb]{revtex4}

\usepackage{graphicx}
\usepackage{dcolumn}
\usepackage{bm}

\begin{document}


\title{Thermodynamic Limit of a Nonequilibrium Steady-State: Maxwell-Type Construction for a Bistable Biochemical System}

\author{Hao Ge$^1$}
\email{gehao@fudan.edu.cn}
\author{Hong Qian$^{2,1,}$}%
\email{qian@amath.washington.edu} \affiliation{$^1$School of
Mathematical Sciences and Centre for Computational Systems Biology,
Fudan University, Shanghai 200433, PRC.
$^2$Department of Applied Mathematics, University of Washington,
Seattle, WA 98195, USA}

\date{\today}

\begin{abstract}
We show that the thermodynamic limit of a bistable
phosphorylation-dephosphorylation cycle has a selection rule for the
``more stable'' macroscopic steady state.  The analysis is akin to
the Maxwell construction.  Based on the chemical master equation
approach, it is shown that, except at a critical point, bistability
disappears in the stochastic model when fluctuation is sufficiently
low but unneglectable.  Onsager's Gaussian fluctuation theory
applies to the unique macroscopic steady state. With initial state
in the basin of attraction of the ``less stable'' steady state, the
deterministic dynamics obtained by the Law of Mass Action is a
metastable phenomenon. Stability and robustness in cell biology are
emergent stochastic concepts.
\end{abstract}

\maketitle

    The statistical physics of a living cell requires a theory
for open molecular system with chemical driving force and free
energy dissipation \cite{ener_diss}. Such system is capable of
reaching a self-organizing state to which many biological functions
are attributed.  The state has been widely known as {\em
nonequilibrium steady-state} (NESS) following M. Klein's concise
terminology \cite{klein_55}.  Two of the most exciting recent
developments in statistical physics are concerned precisely with the
NESS: The fluctuation theorem studies the novel entropy production
characteristics of a NESS \cite{evans_02}; and the one-dimensional
exclusion process deals with highly non-trivial phase behavior
\cite{derrida_98}.

    However, the concept of NESS requires further refinements.
This is the objective of this letter.  From a statistical mechanics
perspective, a NESS is a fluctuating, stochastically stationary
process. It has a stationary probability distribution as well as
correlation functions \cite{lebowitz_55_jqq}.  For a wide class of
physical and biological systems, this state is unique
\cite{footnote1}.  However, from a macroscopic perspective, an open,
driven system can have multiple steady states.  In fact, the
dynamics can be even more complex that include oscillations, chaotic
dynamics, and spatial-temporal chaos \cite{soc}.

    ``Macroscopic'' studies of living cell biochemistry are usually based on
deterministic nonlinear differential equations according to the Law
of Mass Action \cite{fall_book_02}. Currently, it is generally
accepted that a bistability in the deterministic dynamics
corresponds to a bimodal probability density function in the
stochastic approach \cite{HL84,QH09PRSI}. With increasing size of a
chemical reaction systems \cite{footnote2}, there is a separation of
time scale: The transition rates between the two ``macroscopic''
states become infinitesimal $\sim e^{-\alpha V}$, where $V$ is the
systems volume and $\alpha$ is a positive constant. See
\cite{QH09PRSI} for a detailed exposition.

    On the other hand, there is a well-developed, phenomenological
fluctuation theory of NESS in statistical physics, pioneered by
Onsager and Machlup, Lax, and Keizer, among many others
\cite{fluc_NESS}.  One of the most important conclusions from this
classic NESS fluctuation theory is a multivariate Gaussian
fluctuations around a NESS.  This result essentially conforms with
Einstein's equilibrium fluctuation theory.

    In this letter, we shall use a concrete example to provide
insights into this seeming paradox between the current view of NESS
fluctuation, with multiple macroscopic steady states, and the
classic Einstein-Onsager-Lax-Keizer (EOLK) Gaussian theory. Using
the chemical master equation (CME) as the tool and a bistable system
from current biochemical literature, we show that as
$V\rightarrow\infty$, a Maxwell-like construction is necessary. Such
a construction effectively singles out one unique state (or an
attractor in the case of more complex dynamics) in the
thermodynamics limit.  It is around this state that the EOLK theory
applies.  Our analysis confirms the claim that the information
necessary for the Maxwell-type construction is {\em not} present in
the deterministic differential equation model of the system
\cite{NL}; one requires to build a mesoscopic, mechanistic model
with stochasticity in order to gain the required information.  Our
conclusion is that, when dealing with biochemical reaction systems,
one needs to differentiate the thermodynamic limit of a mesoscopic
system and the differential equations based on the Law of Mass
Action. The later follows the Kurtz's theorem \cite{Kur,qian_book};
the thermodynamics limit, however, has to augment the Maxwell
construction, based on which the EOLK fluctuation theory applies.

    Several papers have addressed related issues in the past.
We choose to revisit this important and fundamental issues in NESS
due to the recent resurgent interests in the CME and its
applications to cellular biochemistry.  In addition to \cite{NL},
Keizer developed a Maxwell-type constructions for multiple
nonequilibrium steady states \cite{Kei}.  While the present paper
shares a similar idea, the previous approach was based on the
diffusion approximation to a CME, an approach that can fail to
represent correctly the mesoscopic steady-state \cite{disCMEDiff},
now known as Keizer's paradox \cite{QH09PRSI}.

The analysis performed here is for a particular example of a
biochemical cycle, but what we show is general for nonlinear,
driven chemical reaction systems with multistability.

\vskip 0.3cm

{\bf\em Phosphorylation-dephosphorylation cycle and the CME.}
Biochemical information processing inside cells uses a canonical
reaction system called phosphorylation-dephosphorylation cycle
(PdPC) \cite{qian_book}.  We consider the Ferrell's kinetic model
for PdPC \cite{ferrell_01}, which includes a positive feedback step,
and its reversible extention first studied in \cite{QH05PRL}:
\begin{eqnarray}
&&
E+ATP+K^*\overset{a_1}{\underset{a_{-1}}{\rightleftharpoons}}E^*+ADP+K^*,
\nonumber\\
&& K+2E^*\overset{a_3}{\underset{a_{-3}}{\rightleftharpoons}}K^*,
    \ \ E^*+P\overset{a_2}{\underset{a_{-2}}{\rightleftharpoons}}E+P_i+P,
\label{the_sys}
\end{eqnarray}
in which $E$ and $E^*$ are inactive and active form of a signaling
protein. $K$ and $P$ are enzymes, kinase and phosphatase, that
catalyze the phosphorylation and dephosphorylation respectively.
$K$ and $K^*$ are active and inactive forms of the kinase.  The
chemical reaction of ATP hydrolysis $ATP \rightleftharpoons ADP+Pi$
provides the chemical driving force of the reaction.  In fact the
free energy from the reaction is $\Delta G = k_BT\ln
\left\{a_1a_2[ATP]/(a_{-1}a_{-2}[ADP][Pi])\right\}$. In a cell, $K$
and $P$, ATP,
ADP, and Pi are all at constant concentration, and
$[E]+[E^*]=e_{tot}$. $[z]$ denotes the concentration of the species
$z$.

    The system in (\ref{the_sys}) exhibits bistability according
to a deterministic analysis based on the Law of Mass Action
\cite{ferrell_01}.  It has been further shown that the bistability
is distinctly a driven phenomenon that requires a sufficient large
free energy dissipation \cite{QH05PRL}.  Here we consider its
mesoscopic stochastic model in terms of the CME \cite{qian_book}.

We shall denote $k_1=a_1a_3[ATP]/a_{-3}$,
$k_{-1}=a_{-1}a_3[ADP]/a_{-3}$, $k_2=a_2[P]$ and
$k_{-2}=a_{-2}[P_i][P]$.  Following the previous treatment
\cite{ferrell_01,QH05PRL,qian_book}, we assume the reversible
binding $K+2E^*\rightleftharpoons K^*$ is rapid. The model thus is
simplified into: $E \rightleftharpoons E^*$ with forward and
backward rates $R^{+}(x) =(k_1[K]x^2+k_{-2})(e_{tot}-x)$,
$R^{-}(x)=(k_2+k_{-1}[K]x^2)x$,
and $x(t)=[E^*](t)$.  The energy parameter from ATP hydrolysis
$\gamma=\exp(\Delta G/k_BT)$ $=k_1k_2/(k_{-1}k_{-2})$. $\gamma=1$ is
equivalent to a non-driven system which reaches a unique equilibrium
steady state.  In fact, the equilibrium probability distribution for
the number of $E^*$ is binomial.

The deterministic kinetic model based on the Law of Mass Action is
\begin{eqnarray}
        \frac{dx}{dt}&=&R^{+}(x)-R^{-}(x) = r(x;\theta,\epsilon) \\
        &=&k_2\left\{\theta x^2\left[(e_{tot}-x)- \epsilon x\right]+
            \left[\mu(e_{tot}-x)-x\right]\right\},\nonumber
\label{the_ode}
\end{eqnarray}
in which the three parameters $\theta=k_1[K]/k_2$ represents the
ratio of the activity of the kinase to that of the phosphatase;
$\epsilon=k_{-1}/k_1$ represents the ADP to ATP concentration ratio,
and $\mu=k_{-2}/k_2$ represents the strength of phosphorolysis. In a
living cell, both $\mu$ and $\epsilon$ are small; hence
$\gamma=1/(\mu\epsilon) \gg 1$.

\begin{figure}[ht]
  \begin{center}
   \includegraphics[width=2in,angle=270]{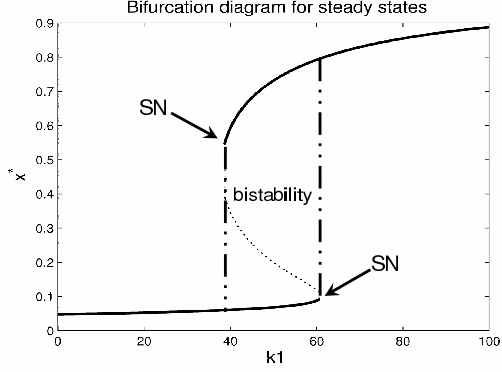}
   \caption{Bifurcation diagram of the steady states of
reaction system in (\ref{the_sys}) and Eq. (\ref{the_ode}), $x^*$,
as a function of the parameter $k_1$. SN denotes saddle-node
bifurcation. Other parameters used in the calculation: $[K]=1$,
$e_{tot}=1$, $k_{-1}=0.01$, $k_2=10$, $k_{-2}=0.5$.} \label{fig1}
  \end{center}
\end{figure}

The fixed points of the Eq. (\ref{the_ode}) are the solution to
$R^+(x)=R^-(x)$. Their stability are determined by the
$\frac{d}{dx}(R^+(x)-R^-(x))$. For some parameter ranges, the Eq.
(\ref{the_ode}) exhibits saddle-node bifurcations
\cite{ferrell_01,QH05PRL}, as shown in Fig. \ref{fig1}.  One obtains
the parameter region for the bistability from simultaneously solving
$r(z)=0$ and $\frac{dr(z)}{dz}=0$, which gives the boundary of the
region of bistability, with a cusp, in $(\theta,\epsilon)$ space (in
terms of $z$ as a parametric curve):
\begin{equation}
    \theta = \frac{2(1+\mu)}{ze_{tot}}-\frac{3\mu}{z^2}, \ \ \
         \epsilon=\frac{2\mu e_{tot}^2-(\mu+1)ze_{tot}}{3\mu e_{tot}z-2(\mu+1)z^2}-1.
\end{equation}

For the stochastic model in terms of the CME, one is interested in
the number of $E^*$, $X$, rather than its concentration. $X$ takes
non-negative integer values and is related to $x=X/V$ where $V$ is
the system's volume.  While $X(t)$ is stochastic, its probability,
$P(X,t)$ satisfies the CME \cite{qian_book}:

$\frac{\partial P(X,t)}{\partial t}
    = VR^{+}\left(\frac{X-1}{V}\right)P(X-1,t)
        +VR^{-}\left(\frac{X+1}{V}\right)$

$\times P(X+1,t)
    -V\left[ R^{+}\left(\frac{X}{V}\right)+R^{-}\left(\frac{X}{V}\right)
        \right]P(X,t)$.

\noindent
Its stationary solution gives the probability distribution in the
NESS:
\begin{equation}\label{exactdis}
    P^{ss}(X)=P_s(0)\prod_{i=1}^X\frac{R^{+}\left(\frac{i-1}{V}\right)}
        {R^{-}\left(\frac{i}{V}\right)}.
\end{equation}

\vskip 0,3cm

{\bf\em Maxwell-type construction and stochastic bifurcation.} When
$V$ is large, by the Euler-MacLaurin summation formula
\begin{equation}
\label{the_ss}
    P^{ss}(xV)\propto A e^{-V\phi(x)},
\end{equation}
where
\begin{eqnarray}
    \phi(x)&=&-\int^x\log
        \left(\frac{R^{+}(y)}{R^{-}(y)}\right) dy\nonumber\\
        &=&e_{tot}\ln (e_{tot}-x)-x\ln\left[\frac{(e_{tot}-x)(\theta x^2+\mu)}
        {x(\theta\epsilon x^2+1)}\right]\nonumber\\
        &&+2\sqrt{\frac{\mu}{\theta}}\arctan
    \left(\sqrt{\frac{\theta}{\mu}}x\right)
        -\frac{2}{\sqrt{\theta\epsilon}} \arctan
        {\sqrt{\theta\epsilon}x}.\nonumber\\
\end{eqnarray}
We note that
\begin{equation}
    \frac{d\phi(x)}{dx} = -\log(R^{+}(x)/R^{-}(x))=-\ln\frac{(e_{tot}-x)(\theta x^2+\mu)}
            {x(\theta\epsilon x^2+1)},\nonumber
\end{equation}
Hence the two stable fixed points of Eq. (\ref{the_ode}) correspond
to the two minima of $\phi(x)$, and the unstable fixed point
corresponds to a maximum. In fact, for each steady state $x^*$,
\begin{equation}
    \phi^{''}(x^*)=\frac{1}{x^*}\left[
    \frac{d\log\frac{R^{-}(x)}{R^{+}(x)}}
    {d\log x}\right]_{x=x^*},
\label{phi_2_p}
\end{equation}
which has the same sign as $d(R^--R^+)(x^*)/dx$. $x^*$ is stable if
$\phi^{''}(x^*)>0$, and unstable otherwise. Near a stable $x^*$
\begin{equation}
    \phi(x) = \phi(x^*)
            + \frac{\phi''(x^*)}{2}\left(x-x^*\right)^2 + \cdots.
\end{equation}
The Gaussian variance of $P^{ss}(x)$ is $(V\phi''(x^*))^{-1}$ which
tends to zero when $V$ tends to infinity.

    The square bracket term in (\ref{phi_2_p}) is called elasticity
\cite{Paulsson} due to its analogue to classical mechanics.
Near the ``more stable'' stable fixed point of
Eq. (\ref{the_ode}), for system with large $V$, a Gaussian, linear
approximation is warranted.  This is the classic theory of EOLK
\cite{fluc_NESS}. The key insight of this theory is the so-called
the fluctuation-dissipation theorem for the NESS, a consequence of
the Markovian Gaussian process.  It provides a relationship among
the linear relaxation kinetic matrix, the noise amplitude, and the
covariance matrix of the Gaussian process. In a similar spirit, Berg
{\em et al.} have put forward a linear noise approximation
\cite{Paulsson}. \cite{Rubi2007} has further illustrated that
Gaussian characteristics is not necessarily only related to
equilibrium fluctuations. Rather, it is determined by linear
dynamics near a steady state \cite{qian01ao05}.

\begin{figure*}[ht]
  \begin{center}
   \includegraphics[width=3in,angle=270]{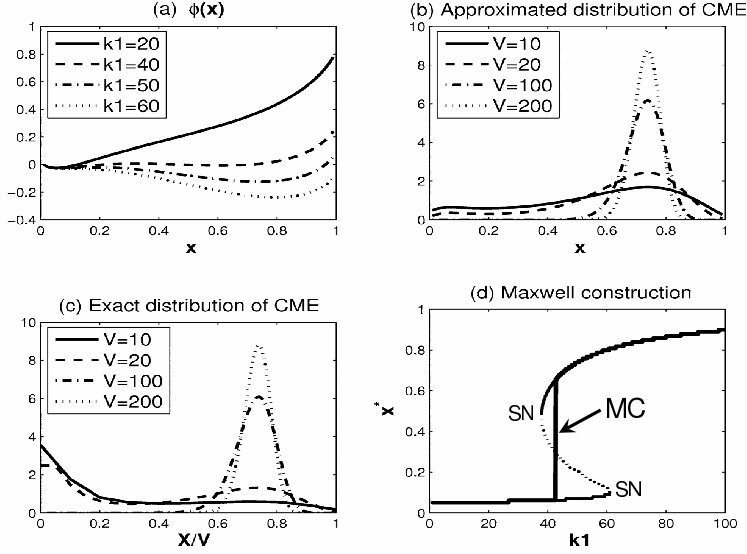}
   \caption{Stationary distribution of CME and Maxwell construction. (a) The shape of $\phi(x)$ varies with the parameter $k_1$;
   (b) Approximated stationary distribution of CME according to Eq. \ref{the_ss} varies with the volume with parameter value $k_1=50$;
   (c) Exact stationary distribution of CME according to Eq. \ref{exactdis} varies with the volume with parameter value $k_1=50$;
   (d) The saddle-node(SN) bifurcation diagram for steady states and the Maxwell construction(MC). Other parameters used in the
calculation: $[K]=1$, $e_{tot}=1$, $k_{-1}=0.01$, $k_2=10$,
$k_{-2}=0.5$.} \label{fig2}
  \end{center}
\end{figure*}

    To our current discussion, the most important feature of
Eq. (\ref{the_ss}) is that the function $\phi(x)$ is independent of
$V$, provided that $V$ is sufficiently large.   Therefore, even
though $\phi(x)$ exhibits bistability which corresponds closely to
the Eq. (\ref{the_ode}), when $V\rightarrow\infty$, only one of the
two stable fixed points is relevant in the thermodynamic limit, and
it is the one with smaller $\phi(x)$. A Maxwell-like construction,
therefore, is necessary at the critical $k_1$  when
$\phi(x^*_1)=\phi(x^*_2)$.  See Fig. \ref{fig2}.

It should be emphasized that bistability in the CME is really a
nonequilibrium phenomenon.  The metastability, however,
can exist even when the white noise is ``additive''.
For example, a diffusive particle restricted in
a bistable potential with vanishing diffusivity \cite{Roy93}.

\vskip 0.3cm

{\bf\em Discontinuity of stochastic entropy production and
first-order phase transition.}  A biochemical NESS is sustained by a
continuous input of chemical energy which is converted to dissipated
heat: Entropy is produced in the process.  The entropy production
rate (epr) can be computed \cite{QH09PRSI}:
\begin{eqnarray}
    \frac{epr}{V}&=&\xi_1p_1^{ss}+\xi_2p_2^{ss}\nonumber\\
                 &=&\triangle
                 G(J_1^{ss}p_1^{ss}+J_2^{ss}p_2^{ss})\nonumber\\
    &\approx& \frac{\triangle G}{\int_{0}^{\infty}e^{-V\phi(x)}dx}
    \left(J_1^{ss}\int_{x^*_1-\epsilon}^{x^*_1+\epsilon}e^{-V\phi(x)}dx\right.\nonumber\\
    &&\left.+J_2^{ss}\int_{x^*_2-\epsilon}^{x^*_2+\epsilon}e^{-V\phi(x)}dx \right).
\end{eqnarray}

The expression here is only valid when the reaction is sufficiently
fast compared to diffusion, so that the reaction rate is only depend
on time and not on the reaction coordinate.

When $V$ tends to infinity, $epr/V$ tends to $\triangle G J_1^{ss}$
if $\phi(x^*_1)<\phi(x^*_2)$, and tends to $\triangle G J_2^{ss}$ if
$\phi(x^*_2)<\phi(x^*_1)$.  Since $J_1^{ss}\neq J_2^{ss}$, the
$epr/V$ is discontinuous at the critical situation when
$\phi(x_1^*)=\phi(x_2^*)$.

In classical equilibrium phase transition theory, a first-order
phase transition has a discontinuity in the first derivative of the
free energy, and a second-order phase transition has a discontinuity
in the second derivative.  According to this classification, the
present (nonequilibrium) phase transition can be considered as first
order.

\vskip 0.3cm

 {\bf\em Summary.} A state of a biological cell, called a
{\em functional cellular attractor} \cite{QH09PRSI}, should be
dynamically stable against various minor purturbations which are
inevitable in living systems.  Thus, it is often thought that
``noise'' added to the biological models only provides moderate
refinements to the behaviors otherwise predicted by the classical,
deterministic description. The present letter, however, shows
something deeper: The relative stability and robustness of the
phosphorylation-dephosphorylation module can {\em not} be properly
inferred without an explicit consideration of the intrinsic noise in
the model.  In cellular biology, it is incorrect to model biological
stability and robustness in terms of deterministic trajectories or
sizes of basins of attractors from a deterministic model.  {\em
Biological stability and robustness are stochastic concepts.} Hence
the presence of noise not only leads to corrections to the
deterministic analysis but may give rise to emergent behaviors.

The CME has now been recognized as a fundamental mathematical theory
for mesoscopic chemical and biochemical reaction systems in a small,
spatially homogeneous volume \cite{qian_book}. Its large volume
limit recovers the Law of Mass Action kinetics \cite{Kur}. However
the deterministic differential equations, while define various
attractors, provide no information on the relative probabilities
between them \cite{NL}. Furthermore, in the thermodynamic limit only
one of the attractors will be dominant with probability 1. The
Maxwell-type construction, thus, enters the CME and becomes an
integral part of a more complete theory. The biochemically
interesting emergent dynamics from a CME, thus, is not the
deterministic differential equations, but rather a stochastic jump
process within a set of discrete states defined by the deterministic
attractors.   This is distinctly a mesoscopic \cite{Laughlin} driven
system phenomenon: When the volume is too large, the time of escaping an
attractor is practically infinite.  Thus, the complex dynamics
disappears. When there is no chemical driving force, i.e.,
$\gamma=1$, the multistability disappears \cite{QH09PRSI}. Near a
given attractor which is a deterministic fixed point, the EOLK
phenomenological Gaussian fluctuation theory applies \cite{fluc_NESS}.
Furthermore, macrosccopic driven system in NESS can behave like an
equilibrium system with a (non-gradient) potential \cite{wang_ao}.


\begin{thebibliography}{99}
\bibitem{footnote1}
Mathematically in terms of the theory of Markov processes, the
condition for unique stationary process is irreducibility.
Ergodicity then naturally follows.
\bibitem{footnote2}
This can be experimentally accomplished by increasing the volume of
a chemical reaction system while keep the concentration invariant.
\end{thebibliography}
\end{document}